\documentclass[a4paper,12pt]{article}
\usepackage{amsfonts,epsfig,latexsym}
\usepackage{pstricks,pst-node}
\catcode`\@=11 \@addtoreset{equation}{section}
\def\theequation{\arabic{equation}}
\def\theequation{\thesection\arabic{equation}}
\newcommand{\nsect}{\setcounter{equation}{0}
\def\theequation{\thesection\arabic{equation}}\section}

\newcommand{\Tr}{{\rm Tr \,}}

\newcommand{\be}{\begin{equation}}
\newcommand{\ee}{\end{equation}}
\newcommand{\ben}{\begin{displaymath}}
\newcommand{\een}{\end{displaymath}}
\newcommand{\ba}{\begin{eqnarray}}
\newcommand{\ea}{\end{eqnarray}}

\newcommand{\bean}{\begin{eqnarray*}}
\newcommand{\eean}{\end{eqnarray*}}

\def\@normalsize{\@setsize\normalsize{15pt}\xiipt\@xiipt
\abovedisplayskip 14pt plus3pt minus3pt%
\belowdisplayskip \abovedisplayskip
\abovedisplayshortskip  \z@ plus3pt%
\belowdisplayshortskip  7pt plus3.5pt minus0pt}
\def\small{\@setsize\small{13.6pt}\xipt\@xipt
\abovedisplayskip 13pt plus3pt minus3pt%
\belowdisplayskip \abovedisplayskip
\abovedisplayshortskip  \z@ plus3pt%
\belowdisplayshortskip  7pt plus3.5pt minus0pt
\def\@listi{\parsep 4.5pt plus 2pt minus 1pt
\itemsep \parsep \topsep 9pt plus 3pt minus 3pt}}
\def\underline#1{\relax\ifmmode\@@underline#1\else
$\@@underline{\hbox{#1}}$\relax\fi} \@twosidetrue \relax
\catcode`@=12
\catcode`\@=11

\def\section{\@startsection{section}{1}{\z@}{3.5ex plus 1ex minus
.2ex}{2.3ex plus .2ex}{\large\bf}}
\def\thesection{\arabic{section}.}

\def\ps@headings{\def\@oddfoot{}\def\@evenfoot{}
\def\@oddhead{\hbox{}\hfill
\makebox[.5\textwidth]{\raggedright\ignorespaces --\thepage{}--
\hfill }}
\def\@evenhead{\@oddhead}
\def\subsectionmark##1{\markboth{##1}{}} }
\ps@headings \catcode`\@=12 \relax

\def\figcap{\section*{Figure Captions\markboth
{FIGURECAPTIONS}{FIGURECAPTIONS}}\list {Fig.
\arabic{enumi}:\hfill}{\settowidth\labelwidth{Fig. 999:}
\leftmargin\labelwidth\advance\leftmargin\labelsep\usecounter{enumi}}}
\relax
\def\tablecap{\section*{Table Captions\markboth
{TABLECAPTIONS}{TABLECAPTIONS}}\list {Table
\arabic{enumi}:\hfill}{\settowidth\labelwidth{Table 999:}
\leftmargin\labelwidth
\advance\leftmargin\labelsep\usecounter{enumi}}}
 \relax
\def\reflist{\section*{References\markboth
{REFLIST}{REFLIST}}\list
{[\arabic{enumi}]\hfill}{\settowidth\labelwidth{[999]}
\leftmargin\labelwidth
\advance\leftmargin\labelsep\usecounter{enumi}}}
 \relax
\catcode`\@=11
\def\marginnote#1{}
\newcount\hour
\newcount\minute
\newtoks\amorpm
\hour=\time\divide\hour by60 \minute=\time{\multiply\hour by60
\global\advance\minute by- \hour}
\edef\standardtime{{\ifnum\hour<12 \global\amorpm={am}%
\else\global\amorpm={pm}\advance\hour by-12 \fi \ifnum\hour=0
\hour=12 \fi
\number\hour:\ifnum\minute<100\fi\number\minute\the\amorpm}}
\edef\militarytime{\number\hour:\ifnum\minute<100\fi\number\minute}
\def\draftlabel#1{{\@bsphack\if@filesw {\let\thepage\relax
\xdef\@gtempa{\write\@auxout{\string
\newlabel{#1}{{\@currentlabel}{\thepage}}}}}\@gtempa
\if@nobreak \ifvmode\nobreak\fi\fi\fi\@esphack}
\gdef\@eqnlabel{#1}}
\def\@eqnlabel{}
\def\@vacuum{}
\def\draftmarginnote#1{\marginpar{\raggedright\scriptsize\tt#1}}
\def\draft{\oddsidemargin -.5truein
\def\@oddfoot{\sl preliminary draft \hfil
\rm\thepage\hfil\sl\today\quad\militarytime}
\let\@evenfoot\@oddfoot \overfullrule 3pt
\let\label=\draftlabel
\let\marginnote=\draftmarginnote
\def\@eqnnum{(\theequation)\rlap{\kern\marginparsep\tt\@eqnlabel}%
\global\let\@eqnlabel\@vacuum}  }
\def\preprint{\twocolumn\sloppy\flushbottom\parindent 1em
\leftmargini 2em\leftmarginv .5em\leftmarginvi .5em \oddsidemargin
-.5in    \evensidemargin -.5in \columnsep 15mm \footheight 0pt
\textwidth 250mmin      \topmargin  -.4in \headheight 12pt
\topskip .4in \textheight 175mm \footskip 0pt
\def\@oddhead{\thepage\hfil\addtocounter{page}{1}\thepage}
\let\@evenhead\@oddhead \def\@oddfoot{} \def\@evenfoot{}  }
\def\titlepage{\@restonecolfalse\if@twocolumn\@restonecoltrue\onecolumn
\else \newpage \fi \thispagestyle{empty}\c@page\z@
\def\thefootnote{\fnsymbol{footnote}} }
\def\endtitlepage{\if@restonecol\twocolumn \else  \fi
\def\thefootnote{\arabic{footnote}}
\setcounter{footnote}{0}}  
\catcode`@=12 \relax

\def\ps@headings{\def\@oddfoot{}\def\@evenfoot{}
\def\@oddhead{\hbox{}\hfill
\makebox[.5\textwidth]{\raggedright\ignorespaces --\thepage{}--
\hfill }}
\def\@evenhead{\@oddhead}
\def\subsectionmark##1{\markboth{##1}{}} }
\ps@headings \relax\def\firstpage#1#2#3#4#5#6{
\begin{document}
\begin{titlepage}
\nopagebreak
\title{\begin{flushright}
\vspace*{-1.8in}
{\normalsize CPTH RR-018.0305}\\[-6mm]
{\normalsize LPT-ORSAY 05/20}\\[-6mm]
{\normalsize UAB-FT-580}\\[-6mm]
{\normalsize hep-th/0503157}\\[-6mm]
\end{flushright}
\vfill {#3}}
\author{\large #4 \\[1.0cm] #5}
\maketitle \vskip -7mm \nopagebreak
\begin{abstract} {\noindent #6}
\end{abstract}
\vfill
\begin{flushleft}
\rule{16.1cm}{0.2mm}\\[-3mm]
$^{\dagger}${\small Unit{\'e} mixte du CNRS et de l'EP, UMR 7644.}\\[-3mm]
$^{\ddagger}${\small Unit{\'e} mixte du CNRS, UMR 8627.}\\
\end{flushleft}
\thispagestyle{empty}
\end{titlepage}}
\def\simlt{\stackrel{<}{{}_\sim}}
\def\simgt{\stackrel{>}{{}_\sim}}
\newcommand{\dal}{\raisebox{0.085cm} {\fbox{\rule{0cm}{0.07cm}\,}}}
\newcommand{\dt}{\partial_{\langle T\rangle}}
\newcommand{\dtbar}{\partial_{\langle\overline{T}\rangle}}
\newcommand{\al}{\alpha^{\prime}}
\newcommand{\mst}{M_{\scriptscriptstyle \!S}}
\newcommand{\mpl}{M_{\scriptscriptstyle \!P}}
\newcommand{\dv}{\int{\rm d}^4x\sqrt{g}}
\newcommand{\lv}{\left\langle}
\newcommand{\rv}{\right\rangle}
\newcommand{\ph}{\varphi}
\newcommand{\abar}{\overline{a}}
\newcommand{\sbar}{\,\overline{\! S}}
\newcommand{\xbar}{\,\overline{\! X}}
\newcommand{\fbar}{\,\overline{\! F}}
\newcommand{\zbar}{\overline{z}}
\newcommand{\dbar}{\,\overline{\!\partial}}
\newcommand{\tbar}{\overline{T}}
\newcommand{\taubar}{\overline{\tau}}
\newcommand{\ubar}{\overline{U}}
\newcommand{\ybar}{\overline{Y}}
\newcommand{\phb}{\overline{\varphi}}
\newcommand{\cm}{Commun.\ Math.\ Phys.~}
\newcommand{\prl}{Phys.\ Rev.\ Lett.~}
\newcommand{\pr}{Phys.\ Rev.\ D~}
\newcommand{\pl}{Phys.\ Lett.\ B~}
\newcommand{\ibar}{\overline{\imath}}
\newcommand{\jbar}{\overline{\jmath}}
\newcommand{\np}{Nucl.\ Phys.\ B~}
\newcommand{\F}{{\cal F}}
\renewcommand{\L}{{\cal L}}
\newcommand{\A}{{\cal A}}
\newcommand{\e}{{\rm e}}
\newcommand{\dslash}{{\not\!\partial}}
\newcommand{\gsi}{\,\raisebox{-0.13cm}{$\stackrel{\textstyle
>}{\textstyle\sim}$}\,}
\newcommand{\lsi}{\,\raisebox{-0.13cm}{$\stackrel{\textstyle
<}{\textstyle\sim}$}\,} \date{}

\firstpage{3118}{IC/95/34} {\Large\bf Five-dimensional Massive Vector
Fields\\ and Radion Stabilization } { Emilian Dudas~$^{\,a,b}$
and Mariano Quiros~$^{\,c}$} {\\[-3mm] \normalsize\sl $^a$ Centre de
Physique Th{\'e}orique$^\dagger$, Ecole Polytechnique, F-91128
Palaiseau, France\\[-2mm] \normalsize\sl $^b$ LPT$^\ddagger$,
B{\^a}t. 210, Univ. de Paris-Sud, F-91405 Orsay, France\\[-3mm]
\normalsize\sl $^c$ ~Instituci\'o Catalana de Recerca i Estudis
Avan\c{c}ats (ICREA)\\[-3mm]
\normalsize\sl Theory Physics Group,
IFAE/UAB,
E-08193 Bellaterra, Barcelona, Spain \\[-3mm]} {We provide a
description of the five-dimensional Higgs mechanism in supersymmetric
gauge theories compactified on the orbifold $S^1/\mathbb Z_2$ by means
of the $\mathcal N=1$ superfield formalism. Goldstone bosons absorbed
by vector multiplets can come either from hypermultiplets or from
gauge multiplets of opposite parity (Hosotani
mechanism). Supersymmetry is broken by the Scherk-Schwarz
mechanism. In the presence of massive hypermultiplets and gauge
multiplets, with different supersymmetric masses, the radion can be
stabilized with positive (de Sitter) vacuum energy. The masses of
vector and hypermultiplets can be fine-tuned to have zero (Minkowski)
vacuum energy.} \break

\section{Introduction}
\label{introduction}

Some time ago it was shown~\cite{pp} in the context of Scherk-Schwarz
type compactifications~\cite{ss} that massive hypermultiplets in five
dimensions have a contribution to the one-loop vacuum energy such
that, when added to the one-loop gravitational contribution, they do
stabilize the radion field with a resulting negative vacuum energy.
The goal of the present paper is to revisit this mechanism of radion
stabilization in five dimensions in theories where massless and
massive hypermultiplets, as well as massive vector multiplets are
present. As we will prove, if hypers and vectors have different 5D
masses, the vacuum energy has different contributions with different
signs and the resulting vacuum energy can generate a radion
stabilization with positive (de Sitter) or zero (Minkowski) vacuum
energy.

Along the way, we provide a description of the five-dimensional Higgs
mechanism in supersymmetric gauge theories. Long time ago
Fayet~\cite{fayet} presented the Higgs mechanism for ${\cal N}=2$
supersymmetric theories in four dimensions and showed that a vector
multiplet can become massive in two different ways. The first way
corresponds to the Goldstone boson coming from a hypermultiplet and
requires a minimal number of one vector and one hypermultiplet to
start with. It corresponds to spontaneous symmetry breaking when the
scalar component of a hypermultiplet acquires a vacuum expectation
value. The simplest realization is the case of a $U(1)$ gauge theory
with one charged hypermultiplet.  The second way corresponds to the
Goldstone boson coming from a vector multiplet and requires more than
one vector multiplet to begin with, i.e. it can only be realized in
non-abelian gauge theories. We will see that it corresponds to the
Hosotani breaking~\cite{Hosotani}, the simplest case being an $SU(2)$
gauge theory. We provide the uplift of these two mechanisms in five
dimensions in the superfield formalism~\cite{mss,agw,mp}.

In Section 2 we present the Higgs mechanism in a 5D theory
compactified on an orbifold $S^1/\mathbb Z_2$ for a $U(1)$ gauge
field, with a Goldstone coming from a charged hypermultiplet, in a
non-linear realization of the gauge symmetry, more convenient in
describing the physical spectrum. We show that the quadratic part of
the action is already 5D and gauge invariant and contains a natural
generalization of Stueckelberg couplings. In Section 3 we discuss the
5D Higgs mechanism for a non-abelian gauge field, in which a $U(1)$
generator of an $SU(2)$ gauge group absorbs a Goldstone coming from
another vector multiplet, again using a convenient non-linear
realization of the gauge symmetry, which parametrizes a vacuum
expectation value for the fifth component of a gauge field (Hosotani
mechanism). In this case, the perturbative lagrangian in the vacuum
spontaneously breaks the 5D Lorentz invariance. In Section 4 we
discuss the appropriate Scherk-Schwarz supersymmetry breaking
deformation, taking into account the different spontaneous breaking of
the $R$-symmetry group in the two cases. In Section 5 we consider
massive hypermultiplets and massive vector multiplets in five
dimensions, of different masses, in the presence of Scherk-Schwarz
supersymmetry breaking.  We show that the one-loop vacuum energy
induced by supersymmetry breaking generates a potential for the radion
which, depending on the number of massless and massive hyper and
vector multiplets and on their masses, can stabilize it with a
positive or zero vacuum energy, for both possible Higgs patterns of
symmetry breaking.  We also provide a four-dimensional
description of the stabilization mechanism in terms of one-loop
corrections to the 4D Kahler potential.
\section{Higgs mechanism in five
dimensions with a Goldstone coming from a hypermultiplet }
\label{hyper}

The framework we are considering is a five dimensional supersymmetric
gauge theory compactified on the interval $S^1/\mathbb Z_2$, leaving
one unbroken supersymmetry in four dimensions. The fifth coordinate
$x^5$ goes from $0$ to $\pi L$.  Let us consider one five dimensional
vector multiplet, described in 4D superfield
language~\cite{mss,agw,mp} by $(V,\Phi)$ with $\mathbb Z_2$ parities
$(+,-)$ and one hypermultiplet $(T_1 , T_2)$ of $\mathbb Z_2$-parities
and $U(1)$-charges $(+,-)$. The invariant action under the $U(1)$
symmetry is~\footnote{We work in what follows in units where the gauge
coupling is one $g_5=1$. An integral over the 5D spacetime is implicit
in the action.}
\begin{eqnarray}
S&=&\int d^2\theta\left[\frac{1}{4}W^{\alpha} W_{\alpha}+
T_1\left(\partial_5-\frac{\Phi}{\sqrt{2}}\right)T_2\right]+{\rm
h.c.}\nonumber\\ &+&\int d^4\theta \left[\left(\partial_5
V-\frac{\Phi+\bar\Phi}{\sqrt{2}}\right)^2+T_1e^V \bar T_1+T_2e^{-V}
\bar T_2 -\xi V \right]
\label{accion1}
\end{eqnarray}
where $\xi$ is a Fayet-Iliopoulos parameter. In fact (\ref{accion1})
is invariant under the gauge transformations
\ba
&&V\to V+\Lambda+\bar\Lambda,\quad \Phi\to \Phi+\sqrt{2}\,\partial_5\,\Lambda
\nonumber\\
&&T_1\to e^{-\Lambda}\,T_1,\quad
 T_2\to e^{\Lambda}\, T_2
\label{gaug1}
\ea
where $\Lambda$ is an $\mathcal N=1$ chiral superfield~\footnote{With
the present field content, the $Z_2$ orbifold induces a gauge anomaly
at the orbifold fixed points. Consistency of the theory can be
achieved in several ways. One possibility is to add a second bulk
charged hypermultiplet such that the surviving fermionic zero mode
cancels the gauge anomaly. If this second hyper gets no vacuum
expectation value, it is a spectator for the analysis of the present
section and counts as an additional hyper in the computation of
Section~5. A second possibility is adding a chiral multiplet localized
on one of the fixed points in order to cancel the global anomaly. In
this case the surviving local anomaly can be cancelled by a 5D
Chern-Simons (CS) term in the bulk Lagrangian $\sim A\wedge F\wedge
F$~\cite{anomalies}. Neither the localized chiral multiplet nor the CS
term are going to alter the analysis of this section. Still a third
possibility is to use a second, bulk odd vector 5D superfield whose
fifth component can shift anomalies in a 5D version of the
Green-Schwarz mechanism~\cite{dgg}.}.

If we want to describe the spontaneous breaking of the $U(1)$ gauge
symmetry when the scalar component of one of the chiral superfields
(say $T_1$) acquires a vacuum expectation value $\sim\sqrt{\xi}$ it is
convenient to change the superfield basis and realize non-linearly the
gauge symmetry by introducing the superfields $Z_{1,2}$ as
\ba
T_1&=&-\sqrt{2}\, m\, e^{-Z_1/\sqrt{2}}\nonumber\\
T_2&=&Z_2\, e^{Z_1/\sqrt{2}}
\ea
where $\xi=2m^2$. In this basis the corresponding gauge
transformations in (\ref{gaug1}) are
\be
Z_1\to Z_1+\sqrt{2}\, \Lambda,\quad Z_2\to Z_2
\label{gauge2}
\ee
and therefore the Goldstone boson is in the scalar component of $Z_1$.
In terms of the new fields, and expanding the action (\ref{accion1})
in a power series of the gauge coupling, the quadratic part of the
action describing the system is given by
\ba
&& S^{(2)}= \int d^2 \theta \left[ {1 \over 4} W^{\alpha} W_{\alpha}
+ m (\Phi - \partial_5 Z_1) Z_2 \right] + {\rm h.c.} \nonumber \\ && +
\int d^4 \theta \left[ \left(\partial_5 V - {\Phi + {\bar \Phi} \over
\sqrt{2}}\right)^2 + {\bar Z}_2 Z_2 + m^2 \left(V - {Z_1 + {\bar Z}_1
\over \sqrt{2}}\right)^2 \right] \
\label{m1}
\ea
where the relationship between $\xi$ and $m$ has been used to cancel
the linear term in $V$. The truncation to the quadratic part of the
action is sufficient for the purpose of finding the physical spectrum,
needed for computing the one-loop vacuum energy in Section 5.

We will now use the notation for the $\mathcal N=1$ field content of the
different superfields: $V\equiv(A_\mu,\lambda,D)$,
$\Phi\equiv(\phi,\psi_\phi,F_\phi)$ and $Z_i\equiv(z_i,\chi_i,F_i)$.
We want to describe the abelian Higgs mechanism in five dimensions
with the imaginary part of $z_1$ being the Goldstone boson absorbed by
the gauge field which becomes massive. We are describing the system
directly in the Stueckelberg, non-linear realization of the gauge
symmetry \cite{stueckelberg}.
The KK gauge boson masses are immediately found out to be
\be
{m_A^{(k)}}^2 \ = \ m^2 + {k^2 \over L^2} \ .
\label{m3}
\ee
The fermion masses are described by the mass terms
\begin{eqnarray} S_{{\rm ferm}} &&=
\sum_k \bigg\{- i m \lambda^{(k)} \chi_1^{(k)} + i {k \over L}
\lambda^{(k)} \psi_{\phi}^{(k)} + {k \over L} \chi_1^{(k)}
\chi_2^{(k)}+ m \psi_{\phi}^{(k)} \chi_2^{(k)} \biggr\}+{\rm h.c.}\nonumber\\
&&\equiv V {\cal
M}^{(KK)} V^T+{\rm h.c.}  \label{m4}
\end{eqnarray}
with $V \equiv (\lambda^{(k)} , \psi_{\phi}^{(k)} , \chi_1^{(k)} ,
\chi_2^{(k)}) $ and ${\cal M}^{(KK)}$ is the fermionic mass
matrix. The eigenvalues of the fermionic mass matrix are
\be ( {\cal
M}^{\dagger} {\cal M} )^{(KK)} \ = \left( m^2 + {k^2 \over L^2}\right) \
{\mathbf 1} \ . \label{m5}
\ee
The canonically normalized Goldstone superfield is ${\hat Z}_1 = m
Z_1$ and the unitary gauge is $Z_1 = 0$. In what follows, for
notational simplicity, we ignore the hat notation on the canonically
normalised Goldstone field.

In order to check the full ${\cal N}=2$ supersymmetry in the spectrum
we are also deriving the scalar mass matrix. By defining the scalar
components
\be \phi \ \equiv \ {\Sigma + i A_5 \over
\sqrt{2}} , \quad z_1 \ \equiv \ {\phi_1 + i a_1 \over \sqrt{2}} \ ,
\label{m6}
\ee
the scalar part of the lagrangian is given by
\ba &&
S_{{\rm scalar}} = |F_1|^2 + |F_2|^2 + |F_{\phi}|^2 + {1 \over 2}
D^2 - \Sigma \partial_5 D - m D \phi_1  \nonumber \\ && + \biggl[
(m \phi - \partial_5 z_1) F_2 + (m F_{\phi} - \partial_5 F_1) z_2 +
{\rm h.c.} \biggr] \ . \label{m7}
\ea
By eliminating the auxiliary fields
\ba && D = m \phi_1 - \partial_5 \Sigma \quad , \quad F_1 = -
\partial_5 {\bar z}_2 \ , \nonumber \\ &&F_2 = - m {\bar \phi} +
\partial_5 {\bar z}_1 \quad , \quad F_{\phi} = - m {\bar z}_2 \
\label{m8}
\ea
and performing some straightforward integrations by parts, we obtain
the scalar potential
\ba && V \ = \ m^2 |z_2|^2 +
|\partial_5 z_2|^2 + {1 \over 2} [m^2 \phi_1^2 + (\partial_5 \phi_1)^2
] \nonumber \\ && + {1 \over 2} [m^2 \Sigma^2 + (\partial_5 \Sigma)^2
] + {1 \over 2} (m A_5 - \partial_5 a_1 )^2 \ . \label{m9}
\ea
By performing now the standard KK expansions
\be a_1 = {1 \over \sqrt{ \pi
L}} \sum_{k=0}^{\infty} r_k \cos{k y \over L} \ a_1^{(k)} \quad ,
\quad A_5 = {1 \over \sqrt{ \pi L}} \sum_{k=1}^{\infty} \sin{ k y
\over L} \ A_5^{(k)} \ , \label{m10}
\ee
where $r_0 = 1/\sqrt{2}$ and $r_k = 1$ for $k = 1 \cdots \infty$
and defining the mass eigenstates
\ba \xi_1^{(k)} \ = \ {1 \over \sqrt{m^2 + k^2/L^2}} \ \left[m
a_1^{(k)} - {k \over L} A_5^{(k)} \right] \qquad , \qquad {\rm for } \
k = 0, \dots, \infty \ , \nonumber \\ \xi_2^{(k)} \ = \ {1 \over
\sqrt{m^2 + k^2/L^2}} \ \left[ {k \over L} a_1^{(k)} + m A_5^{(k)}
\right] \qquad , \qquad {\rm for } \ k = 1, \dots, \infty \ ,
\label{m11}
\ea
we find the mass lagrangian for KK states
\ba &&
S_{\,{\rm scalar}} \ = \ \sum_{k=1}^{\infty} \left(m^2 + {k^2 \over L^2}\right)
\left[ |z_2^{(k)}|^2 + {1 \over 2} (\Sigma^{(k)})^2 + {1 \over 2}
(\xi_2^{(k)})^2 \right] \ \nonumber \\ && + \sum_{k=0}^{\infty}
\left[ {1 \over 2} \left(m^2 + {k^2 \over L^2}\right) (\phi_1^{(k)})^2 + r_k
\sqrt{m^2 + {k^2 \over L^2}} A_{\mu}^{(k)} \partial^{(\mu)}
\xi_1^{(k)} \right] \ . \label{m12}
\ea

The physical interpretation is that $\xi_1^{(k)}$ are the Goldstone
bosons eaten by the KK modes of the gauge fields which become massive,
while five scalar degrees of freadom $z_2^{(k)}, \phi_1^{(k)} ,
\Sigma^{(k)}$ and $\xi_2^{(k)}$ become massive with the same mass as
the fermions (\ref{m5}). The three gauge field degrees of freedom plus
the five massive scalar ones do correctly match the eight massive
fermionic degrees of freedom in (\ref{m5}).

We would like now to check the 5D Lorentz invariance of the effective
action. The fermionic part of the 5D lagrangian is best expressed by
introducing the Dirac fermions
\be
\Psi_1 = (\psi_{\phi} \,, \, -i {\bar \lambda})^T \quad ,
\quad \Psi_2 = (\chi_1\,  , - {\bar \chi}_2 )^T \ . \label{m14}
\ee
Defining as usual the 5D gauge field $A_M = (A_{\mu}, A_5 )$ and
combining the 4D gauge bosons and scalars terms, we find the total
quadratic action
\ba && S^{(2)} \ = \ - {1 \over 4} F_{MN} F^{MN} - |\partial_M
z_2|^2 - m^2 |z_2|^2 - {1 \over 2} [(\partial_M \phi_1)^2+ m^2
\phi_1^2 ]
\nonumber \\
&& - {1 \over 2} [(\partial_M \Sigma)^2 + m^2 \Sigma^2 ]
- {1 \over 2} (m A_M - \partial_M a_1 )^2 \ \nonumber \\
&&  - i {\bar \Psi}_1 \gamma^M \partial_M
\Psi_1 - i {\bar \Psi}_2 \gamma^M \partial_M \Psi_2
- m  ( {\bar \Psi}_1 \Psi_2 + {\bar \Psi}_2 \Psi_1 )
\ . \label{m13}
\ea
Notice the obvious generalisation of the Stueckelberg mixing
between the gauge field $A_M$ and the Goldstone boson $a_1$ which
appears in the last term of the second line of (\ref{m13}).

\section{Higgs mechanism in five
dimensions with a Goldstone coming from a vector multiplet }
\label{vector}

We will now consider the case of an $SU(2)$ gauge theory broken into
$U(1)$ by the orbifold boundary conditions~\cite{myreview}. We then
have three five dimensional vector multiplets, described in 4D
superfield language~\cite{mss,agw,mp} by $(V_a,\Phi_a)$ with $a=1,2,3$
and $\mathbb Z_2$ parities given in the Table.
\begin{center}
\begin{tabular}{||c|c|c||}
\hline\hline
$\mathbb Z_2$--parity & $+$ & $-$\\ \hline Vector & $V_3$
& $V_{1,2}$\\ Chiral & $\Phi_{1,2}$& $\Phi_3$\\
\hline\hline
\end{tabular}
\end{center}
The action of the system can be written as~\cite{mp}
\ba
S&=&\frac{1}{4}\int d^2\theta\, \Tr W^\alpha W_\alpha + {\rm h.c.} \nonumber\\
&+& 2 \int d^4\theta\, \Tr\left[\left\{e^{V/2},\partial_5e^{-V/2}\right\}+
\frac{1}{\sqrt{2}}\left(e^{V/2}\bar\Phi e^{-V/2}+e^{-V/2}\Phi e^{V/2}\right)
\right]^2
\label{accion2}
\ea
where $V=V_a\sigma^a/2$ and $\Phi=\Phi_a\sigma^a/2$.  In fact the
action (\ref{accion2}) is invariant under the supergauge
transformations
\be
V\to e^\Lambda\, e^V\, e^{\bar\Lambda}, \quad \Phi\to e^\Lambda\,(\Phi-\sqrt{2}
\partial_5)\,e^{-\Lambda}
\label{gauge3}
\ee
where $\Lambda=\Lambda_a\sigma^a/2$.

We will now describe the spontaneous breaking of the surviving $U(1)$
gauge symmetry by the Hosotani mechanism. The $\mathcal N=1$ notation
is $V_a=(A_a^\mu,\lambda_a,D_a)$, $\Phi_a=(\phi_a,\psi_a,F_a)$ and
$\phi_a=(\Sigma_a+iA^5_a)/\sqrt{2}$. We want to consider the Higgs
mechanism where the imaginary part of $\phi_1$, $A^5_1$, acquires a
vacuum expectation value. Analogously to the first case, we are
describing the system directly in the Stueckelberg, non-linear
realization of the gauge symmetry by the field redefinition
\be
\phi=\phi_3\sigma^3/2+e^{\chi_1 \sigma^3/2}(i\sqrt{2} \ m \ \sigma^1/2+
\chi_2\sigma^2/2)e^{-\chi_1\sigma^3/2}
\label{phi}
\ee
in such a way that under the surviving gauge transformation
$\Lambda_3$ the fields transform as
\be
\phi_3\to\phi_3-\sqrt{2} \ \partial_5\Lambda_3,\quad
\chi_1\to\chi_1+\Lambda_3,\quad \chi_2\to\chi_2
\ee
and the field $\chi_1$ can be gauged away. In fact the relation
between the parametrization in (\ref{phi}) and the $\phi_a$ basis is
given by
\ba
\phi_1&=&i(\sqrt{2}m\cosh \chi_1-\chi_2\sinh\chi_1)=i\sqrt{2}m+\mathcal
O(\chi_1^2,\chi_1\chi_2)\nonumber\\
\phi_2&=&\chi_2\cosh\chi_1-\sqrt{2}m\sinh\chi_1=\chi_2-\sqrt{2}m\chi_1+\mathcal
O(\chi_1^2,\chi_1\chi_2)
\label{linear}
\ea

The quadratic part of the $D$-term contribution to the action
(\ref{accion2}) can be written as
$$ 2\int d^4\theta\,\Tr\left[\partial_5
V-\frac{\phi+\bar\phi}{\sqrt{2}}+\frac{1}{2 \sqrt{2}}
\left[V,\phi-\bar\phi\right]\right]^2
$$
and in terms of the new fields the action (\ref{accion2}) to quadratic
order is
\ba S^{(2)}&=&\frac{1}{4}\int d^2\theta\, W^\alpha_a W_\alpha^a + {\rm h.c.} \\
&+& \int d^4\theta\, \left\{\left[\partial_5V_1
\right]^2+\left[\partial_5V_3+m
V_2-\frac{\phi_3+\bar\phi_3}{\sqrt{2}}\right]^2+\left[\partial_5V_2-m
V_3-\frac{\phi_2+\bar\phi_2}{\sqrt{2}}\right]^2\right\}\nonumber
\label{accion4}
\ea
where the field $\phi_{2}$ is defined as function of $\chi_{1,2}$ as
in Eq.~(\ref{linear}). The action (\ref{accion4}) is invariant under
the gauge transformations
\ba
&& \delta V_3 = \Lambda_3 + {\bar \Lambda}_3 \quad , \quad
\delta \phi_3 = \sqrt{2} \ \partial_5 \Lambda_3 + \sqrt{2} \ m \Lambda_2 \
, \nonumber \\
&& \delta V_2 = \Lambda_2 + {\bar \Lambda}_2 \quad , \quad
\delta \phi_2 = \sqrt{2} \ \partial_5 \Lambda_2 - \sqrt{2} \ m
\Lambda_3 \ . \label{m17}
\ea
Notice that at the quadratic order there is no gauge invariance
corresponding to the $V_1$ gauge boson, which is therefore a general
real vector superfield. Since it is completely decoupled from the
rest of the lagrangian and will play no role in our discussion, we
ignore it in what follows.

In order to derive the scalar mass matrix, we start from the scalar
part of the action
\be
S_{{\rm scalar}} = {1 \over 2}
(D_2^2+D_3^2)  - (\partial_5 D_3 +  m D_2) \Sigma_3
-  (\partial_5 D_2 -  m D_3)  \Sigma_2 \ . \label{m18}
\ee
By eliminating the auxiliary fields
\ba
&& D_3 = m \Sigma_2 - \partial_5 \Sigma_3  \ , \nonumber \\
&& D_2 = -m \Sigma_3 - \partial_5 \Sigma_2    \  \label{m19}
\ea
and integrating by parts, we find the four-dimensional scalar
potential
\be V \ = \ {1 \over 2} \biggl[ (\partial_5 \Sigma_3)^2 + (\partial_5
\Sigma_2)^2 + m^2 \Sigma_3^2 + m^2
\Sigma_2^2-4m\Sigma_2\partial_5\Sigma_3 \biggr] \ . \label{m20} \ee
The full lagrangian in this case has spontaneously broken 5D Lorentz
invariance due to the vacuum expectation value of the fifth component
of $A_1$, whereas by construction the original lagrangian is clearly
5D Lorentz invariant.  It is instructive, nonetheless, to write the
resulting lagrangian at the quadratic level (by again neglecting the
decoupled $V_1$ superfield) and check the Stueckelberg type couplings
of gauge fields to the Goldstone modes. The result is

\ba
&& S^{(2)} = - {1 \over 4} \left( \left(F^{\mu \nu}_2\right)^2 +
\left(F^{\mu \nu}_3\right)^2\right) - {1 \over 2} \left( F^{\mu 5}_2 +
m A^{\mu}_3\right)^2 - {1 \over 2} \left( F^{\mu 5}_3 - m
A^{\mu}_2\right)^2 \nonumber \\
&& - {1 \over 2} \left[
(\partial_{\mu} \Sigma_2)^2 + (\partial_{\mu} \Sigma_3)^2 \right] - {1
\over 2} ( m \Sigma_2 - {\partial_5 \Sigma_3})^2 - {1 \over 2} ( m
\Sigma_3 + {\partial_5 \Sigma_2})^2 \ \nonumber \\ && - i \sum_{j=2,3}
\left( \lambda_j \sigma^{\mu} {\partial_{\mu} {\bar \lambda}_j} +
\psi_j \sigma^{\mu} {\partial_{\mu} {\bar \psi}_j} \right) + i \left[
\lambda_2 (\partial_5 +m) \psi_3 - \lambda_3 (\partial_5 +m) \psi_2
\right] \ . \label{m21}
\ea
The Stueckelberg couplings between the gauge field $A^{\mu}_2$ and the
Goldstone $A^5_3$ and $A^{\mu}_3$ and the Goldstone $A^{5}_2$ clearly
apppear in the first line of (\ref{m21}).

By performing the standard KK expansion
\be
\Sigma_2 = {1 \over \sqrt{ \pi L}} \sum_{k=0}^{\infty} r_k \cos{k y \over
  L} \ \Sigma_2^{(k)} \quad , \quad
\Sigma_3 = {1 \over \sqrt{ \pi L}} \sum_{k=1}^{\infty} \sin{ k y \over
  L} \ \Sigma_3^{(k)} \ , \label{m22}
\ee
we find the KK mass eigenvalues
\be {m_{\pm}^{(k)}} = \frac{k}{L}\pm m\quad (k\neq 0),\quad m^{(0)}=m
. \label{m23} \ee
The gauge boson and fermion masses of the KK states have a similar
structure to those in (\ref{m23}).

Notice that in the particular case we were studying in detail the
vector multiplets $V_{2,3}$ can be considered (from the point of view
of number of degrees of freedom) as a complex vector multiplet since
the vacuum expectation value of $\phi_1$ gives a mass both to the
unbroken $U(1)$ generated by $\{\sigma^3/2\}$ and the $U(1)$ generated
by $\{\sigma^2/2\}$, already broken by the orbifold boundary
conditions. In other cases the counting is a bit more complicated but
similar. Take for instance the case where $SU(3)\to SU(2)\otimes U(1)$
by the orbifold boundary conditions where the $SU(3)$ generators
$T_{1,2,3,8}$ are even and $T_{4,5,6,7}$ are odd. We can further break
$SU(2)\otimes U(1)$ by the Hosotani mechanism when some of the
imaginary parts of the scalar components of $\phi_{4,5,6,7}$, say
$\phi_4$, acquire a vacuum expectation value. Only the vector fields
along the generators commuting with $T_4$ will not acquire any mass,
namely $V_4$ and $V_3+\sqrt{3}V_8$. The other six real (three complex)
vectors will get a massive spectrum similar to that in (\ref{m23}).

\section{Scherk-Schwarz supersymmetry breaking for the massive vector
multiplet}\label{SS}

The next step is adding a source of soft supersymmetry breaking. The
best suited for our purposes is the generalized dimensional reduction
\`a la Scherk and Schwarz~\cite{ss}, in which various fields are not
exactly periodic in the compact coordinate $x^5$ as in the standard KK
reduction, but periodic only up to a symmetry transformation
$U(\omega)$
\be {\hat \phi} (x^5 + 2 \pi L) \ = \ U (\omega) \ {\hat \phi} (x^5) \
, \label{s1} \ee
where $\omega$ is a quantized parameter in a string theory context.
In an orbifold compactification there are restrictions on the twist
matrix $U$, namely it has to commute \cite{fkpz} with the orbifold action $g$,
$[U,g]=0$. In the particular case of the $S^1/\mathbb Z_2$
compactification, the form of the twist matrix and the consistency
condition to be satisfied are~\cite{dg}
\be
U \ = \ e^{M x^5 } \qquad , \qquad \{ M , g \} \ = \ 0 \ . \label{s2}
\ee

\subsection{The case of the hyper Goldstone multiplet}

In this case the original $SU(2)\otimes U(1)_R$ symmetry is
spontaneously broken to the $U(1)_R$ subgroup~\cite{fayet}, that
contains $R$-parity as a discrete subgroup. Therefore the appropriate
symmetry to be used in the Scherk-Schwarz reduction in this case
is~\cite{mariano} the $R$-parity, under which all fermions from the
vector and the Goldstone hyper and the bosons from the matter
hypermultiplets have odd boundary conditions $U=-1$, whereas the rest
of fields are periodic. This corresponds to the following Kaluza-Klein
fermionic expansion
\ba
&& ( \lambda , \chi_1)  =
{1 \over \sqrt{ \pi L}} \sum_{k=0}^{\infty} r_k \
\cos{(k+1/2) y \over L} \ ( \lambda^{(k)}, \chi_1^{(k)})  \ , \nonumber \\
&& ( \psi_{\phi} , \chi_2 )
\ = \ {1 \over \sqrt{ \pi L}} \sum_{k=1}^{\infty} \ \sin{ (k+1/2) y \over
  L} \ ( \psi_{\phi}^{(k)}, \chi_2^{(k)}) \ , \label{s3}
\ea
whereas the bosonic KK expansion for the vector and Goldstone hyper is
unchanged. The KK bosonic masses are like in (\ref{m3}), (\ref{m5}),
whereas the fermionic masses are shifted according to
\be
{m_f^{(k)}}^2 \ = \ m^2 + {(k+1/2)^2 \over L^2} \ . \label{s4}
\ee

\subsection{The case of the vector Goldstone multiplet}

In this case the original symmetry $SU(2)\otimes U(1)_R$ symmetry is
spontaneously broken to $SU(2)_R$~\cite{fayet}, and the appropriate
twist matrix which satisfies the consistency requirement (\ref{s2}) is
the $U(1)$ subgroup of the $SU(2)_R$ $R$-symmetry in five
dimensions~\cite{dg}, which acts on the fermions of the two vector
multiplets as
\ba
&& \pmatrix{ \hat \lambda_3 \cr \hat \psi_3 } ~\equiv~ ~ \pmatrix {
               \cos (\omega y/L) & -\sin (\omega y/L) \cr
               \sin (\omega y/L) & \cos (\omega y/L) \cr}~
         \pmatrix{ \lambda_3 \cr \psi_3 } \  , \nonumber \\
&& \pmatrix{ \hat \lambda_2 \cr \hat \psi_2 } ~\equiv~ ~ \pmatrix {
               \cos (\omega y/L) & \sin (\omega y/L) \cr
               - \sin (\omega y/L) & \cos (\omega y/L) \cr}~
         \pmatrix{ \lambda_2 \cr \psi_2 } \ , \label{s5}
\ea
where $\omega = 1/2$ in the simplest examples worked out in the
literature. For a general $\omega$, the fermionic spectrum
(eigenvalues of the fermionic mass matrix) is shifted according to
\be
m_{f\,\pm}^{(k)} \ = \frac{k\pm(\omega+mL)}{L}
 \label{s6}
\ee
In the particular case $\omega+mL = 1/2$ the up and the down mass shifts
in (\ref{s6}) are equivalent when summing over the whole KK tower of states.

\nsect{One-loop vacuum energy and radion stabilization}

In this section we will compute the one-loop effective potential in
the background of the radion field. We will consider, on top of the
gravitational contribution a number of massless ($N_v$ vectors and
$N_h$ hypers) and massive ($N_H$ hypers and $N_V$ vectors)
multiplets\footnote{We are using a compact notation in order to cover
both cases of Sections 2 and 3. For the case of the hyper Goldstone of
Section 2, $N_V$ is actually twice the number of the vector
multiplets, whereas $N_V$ is equal to the number of vector multiplets
for the case of vector Goldstone of Section 3.}. We will assume that
supersymmetry is broken by the SS mechanism with an arbitrary
parameter $\omega$ that should be fixed to the value $1/2$ if the
Goldstones come from hypermultiplets.

The formalism of massive vector multiplets was presented in previous
sections. For completeness we will now consider the case of
hypermultiplets with a supersymmetric mass $M$. Let us then consider
the hypermultiplet
$(\chi_1,\chi_2)\equiv(\chi_1,\chi_2;\psi;F_1,F_2)$, with parities
$(+,-)$ and let us introduce the odd mass $M(x^5)=\epsilon(x^5)M$,
where $\epsilon(x^5)$ is the sign function. The supersymmetric
lagrangian is given in $\mathcal N=1$ language by~\cite{agw,mp}
\be \mathcal L=\int
d^4\theta\left[\bar\chi_1\chi_1+\bar\chi_2\chi_2\right]- \left\{\int
d^2\theta \chi_1\left[\partial_5-M(x^5)\right]\chi_2+\
{\rm h.c.}\right\}
\ee
or in components
\begin{eqnarray}
\mathcal L&=&- \left|\partial_M\chi_1\right|^2 - \left|\partial_M\chi_2\right|^2
- i\bar\psi\gamma^M\partial_M\psi+M(x^5)\bar\psi\psi\nonumber\\
&-&\left[M^2+\partial_5 M(x^5)\right]\left|\chi_1\right|^2-
\left[M^2-\partial_5 M(x^5)\right]\left|\chi_2\right|^2
\label{hyperM}
\end{eqnarray}

The mass eigenvalues corresponding to the system described by the
lagrangian (\ref{hyperM}) were computed in
Ref.~\cite{vonGersdorff:2003qf}. In particular the eigenvalues of the
fermionic mass matrix are given by
\be
m_k^2=\frac{k^2}{L^2}+M^2(1-\delta_{k0})
\ee
and those of the bosonic mass matrix by the solution to the equation
\be
\sin^2(\pi\omega)=\frac{m_k^2}{\Omega_k^2}\sin^2(\Omega_k\pi L)
\ee
where $\Omega_k=\sqrt{m_k^2-M^2}$.

Up to this point we have parametrized the theory as a function of the
interval length scale $L$. We will now introduce the physical radion
field $R$. In order to do that we will parametrize the 5D metric in
the Einstein frame in terms of the ``dilaton'' field $\varphi$
as~\cite{Appelquist}
\be
ds^2=\varphi^{-1/3}g_{\mu\nu}dx^\mu dx^\nu+\varphi^{2/3}dx^5 dx^5
\ee
where $x^5$ goes from $0$ to $\pi L$. The radion field whose VEV
determines the size of the extra dimension is the physical radius
given by $R=\varphi^{1/3} L$. The length $L$ is unphysical. It will
drop out once the VEV of the radion $\varphi_c^{1/3}L$ is fixed and
the effective 4D theory will only depend on $R$. Notice that in
previous sections we were identifying $R$ with $L$ in units of
$\varphi_c$. From here on we will turn on the explicit dependence on
$\varphi_c$. Turning on a value of $\varphi_c\neq 1$ in the action
amounts (when defining canonically normalized fields
$\tilde\phi=\varphi_c^{1/6}\phi$) to changing $L\to\varphi_c^{1/2}L$,
$M\to \varphi_c^{-1/6}M$ such that $LM\to\varphi_c^{1/3}LM=RM$. All
the previously considered spectra scale correspondingly with
$\varphi_c$ as we will see now.

We will first consider the massless sector, i.e. the gravitational
sector plus $N_v$ massless vector multiplets and $N_h$ massless
hypermultiplets. The mass squared of KK modes of the gravitino, gauginos and
hyperscalars is given by:
\be
m_k^2=\frac{1}{\varphi_c}\frac{(k+\omega)^2}{L^2}
\label{KKm0}
\ee
and those of the rest of KK modes is given by (\ref{KKm0}) after
fixing $\omega=0$. The Coleman-Weinberg one-loop effective
potential~\cite{CW} can be easily computed to be~\cite{Delgado:1998qr}
\be V_{\rm eff}=\frac{2+N_v-N_h}{64\pi^6 \varphi_c^2 L^4}\left[3
Li_5(r)+3 Li_5(1/r)-6\zeta(5)\right]
\label{massless}
\ee
where $r=\exp(2\pi i\omega)$, the polylogarithms are defined as
\be
Li_n(x)=\sum_{k=1}^\infty \frac{x^k}{k^n}
\ee
and $\zeta(n)=Li_n(1)$.

The effective potential of $N_H$ massive hypermultiplets with a common
supersymmetric mass $M_H$ was computed in
Ref.~\cite{vonGersdorff:2003qf}. In the presence of a SS twist
$\omega$ the effective potential is given by
\be
V_{\rm eff}=\frac{N_H}{64\pi^6\phi_c^2 L^4}g(x_H,\omega)
\ee
where $x_H=M_H\pi R$ and
\be
g(x,\omega)=8\int_0^\infty dz
z^3\log\left[1+\frac{(z^2+x^2)\sin^2(\pi\omega)}
{z^2\sinh^2\left(\sqrt{z^2+x^2}\right)}\right]
\ee

Finally the mass eigenvalues of fermions in gauge multiplets with
supersymmetric mass $M_V$ and where the Goldstone bosons are in
hypermultiplets is given by [Eq.~(\ref{s4})]
\be
m_k^2=\frac{M_V^2}{\varphi_c^{1/3}}+\frac{1}{\varphi_c}
\frac{(k\pm\omega)^2}{L^2}
\ee
and the effective potential is given by
\be
V_{\rm eff}=\frac{N_V}{64\pi^6\varphi_c^2L^4}f(x_V,\omega)
\label{vector1}
\ee
where $x_V=M_V\pi R$ and
\begin{eqnarray}
f(x,\omega)&=&4x^2Li_3\left(r e^{-2x}\right)+6xLi_4\left(r
e^{-2x}\right)+3Li_5\left(r
e^{-2x}\right)+{\rm h.c.}\nonumber\\
&&-8x^2Li_3\left(e^{-2x}\right)-12 x
Li_4\left(e^{-2x}\right)-6Li_5\left(e^{-2x}\right)
\label{f1}
\end{eqnarray}
Notice that here, as we already explained in the previous section,
$\omega=1/2$.

Similarly the mass eigenvalues of fermions in complex gauge multiplets
with supersymmetric mass $M_V$ and where the Goldstone bosons are in
gauge multiplets (Hosotani breaking) is given by [Eq.~(\ref{s6})]
\be
m_k^2=\frac{1}{\varphi_c}\left(\frac{k\pm(\omega+M_VR)}{L}\right)^2
\ee
and its contribution to the effective potential is given by
Eq.~(\ref{vector1}) with the function $f(x,\omega)$ defined by
\be
f(x,\omega)=3 Li_5\left(r\,e^{2ix}\right)-3 Li_5\left(e^{2ix}\right)+{\rm h.c.}
\label{f2}
\ee
As it is clear from (\ref{f2}) the potential from massive gauge bosons
when the gauge symmetry is broken by the Hosotani mechanism is
oscillatory and therefore essentially different to the case where
Goldstone bosons are in hypermultiplets.

We will first refer to the case where Goldstone bosons are in
hypermultiplets and therefore supersymmetry breaking is achieved with
a Scherk-Schwarz parameter $\omega=1/2$.  In
Ref.~\cite{vonGersdorff:2003rq} a radion potential with anti de Sitter
minimum was created by the presence of $N_H$ hypermultiplets with a
common supersymmetric mass. The conditions for such a behaviour,
$V_{\rm eff}\to+\infty$ when $R\to 0^+$ and $V_{\rm eff}\to 0^-$ when
$R\to\infty$, were fulfilled if $2+N_v-N_h-N_H<0,\ 2+N_v-N_h>0$.  We
now want a radion potential exhibiting a minimum with positive (de
Sitter) or zero (Minkowski) vacuum energy. A necessary condition for
that is that the potential $V_{\rm eff}\to+\infty$ when $R\to 0^+$ and
$V_{\rm eff}\to 0^+$ when $R\to\infty$. These conditions are fulfilled
provided that
\be
2+N_v+N_V-N_h-N_H<0,\qquad 2+N_v-N_h<0
\ee
respectively. Then we can write the effective potential as
\be
V_{\rm eff}=\frac{N_h-2-N_v}{64}M^6_HL^2 \mathcal V(x) \ , \label{o01}
\ee
where $x\equiv x_H=\pi M_H R$ and
\be \mathcal V(x)=\frac{1}{x^6}\left\{ -f(0,\omega)+\delta_H
g(x,\omega)+\delta_V f(\alpha x,\omega)\right\}\equiv \frac{1}{x^6} v(x)
\label{potencial}
\ee
We have defined in (\ref{potencial})
\be
\delta_{H,V}=\frac{N_{H,V}}{N_h-2-N_v}
\ee
and $\alpha=M_V/M_H$.

\begin{center}
\begin{figure}[htb]
\centerline{\epsfxsize=3.5in\epsfbox{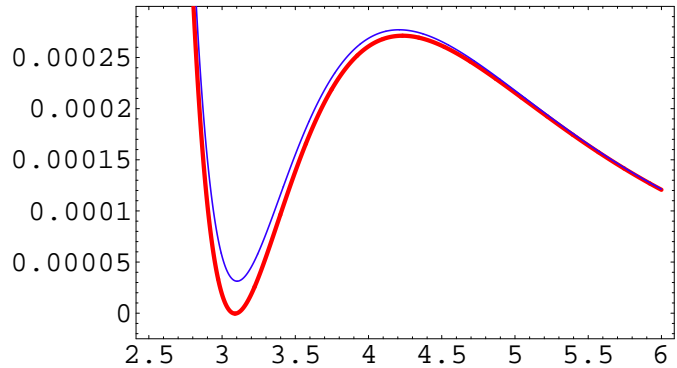}}
\caption{\it Radion effective potential $\mathcal V(x)$ as defined in
the text, with Goldstone bosons in hypermultiplets, for
$\delta_H=\delta_V=1$ and $\omega=1/2$. Upper (blue-thin) potential
has de Sitter minimum: it is for $\alpha=0.27$. Lower (red-thick)
potential with Minkowski minimum: $\alpha\simeq 0.27$ is tuned to zero
vacuum energy. }\label{figura}
\end{figure}
\end{center}

It is easy to see that the minimum condition with zero vacuum energy
$\mathcal V'(x)=\mathcal V(x)=0$ is equivalent to the simplest one
$v'(x)=v(x)=0$ and that a necessary condition for de Sitter or
Minkowski minimum is $M_V<M_H$,~i.e. $\alpha<1$. In fact it is easy to
work out cases where this happens. A simple example is provided in
Fig.~\ref{figura} where we have considered the case
$\delta_H=\delta_V=1$ which corresponds to
$N_H=N_V=2(N_h-N_v-2)$. The upper (thin) curve exhibits a potential
where the minimum is de Sitter: it corresponds to $\alpha=0.27$. In
the lower (thick) curve the value of $\alpha$ has been fine-tuned to
have a zero vacuum energy (Minkowski minimum). The minimum is located
at $x\simeq 3.1$ which corresponds to values of the supersymmetric
masses, $M_H\simeq 1/R$ and $M_V\simeq 0.27/R$. Of course other
similar cases can easily be found.

\begin{center}
\begin{figure}[htb]
\centerline{\epsfxsize=3.5in\epsfbox{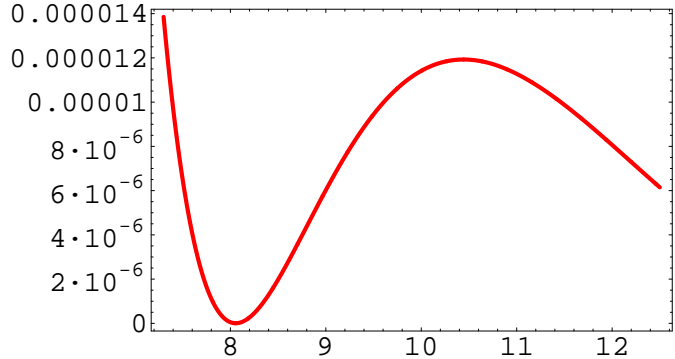}}
\caption{\it Radion effective potential for Goldstone bosons in gauge
multiplets for $2\delta_H=3\delta_V=3$, $\omega=1/2$ and
$\alpha\simeq 0.39$. }\label{figura2}
\end{figure}
\end{center}

In the case where Goldstone bosons are in gauge multiplets, and the
breaking of the gauge symmetry is by the Hosotani mechanism, the
potential from massive gauge bosons is an oscillatory function as
given in Eq.~(\ref{f2}). In this case it is also possible to tune the
parameters such that there is a Minkowski minimum for the function
$v(x)$ in (\ref{potencial}) at $x=x_0$ and other AdS minima for higher
values of $x$. However since the whole effective potential is $\propto
v(x)/x^6$ the Minkowski minimum remains and AdS minima are suppressed
by the factor $1/x^6$.  On the other hand the tunnelling probability
to the AdS minima is exponentially small and the Minskowski vacuum is
essentially stable on cosmological times~\cite{vonGersdorff:2003rq}. A
particular example is presented in Fig.~\ref{figura2} where we have
considered the case $2\delta_H=3\,\delta_V=3$,
i.e.~$2N_H=3N_V=3(N_h-2-N_v)$, a Scherk-Schwarz parameter $\omega=1/2$
and $\alpha\simeq 0.39$. The minimum is located at $x\simeq 8.1$ which
corresponds to values of the supersymmetric masses $M_H\simeq 2.6/R$
and $M_V\simeq 1/R$.

Finally it is also useful to present a four-dimensional description of
the stabilization mechanism in terms of the standard 4D supergravity
lagrangian~\cite{cfgp}. This description is possible for large radii
where one can truncate the higher derivative corrections to the
effective action. The radius $R$ and the fifth component of the gauge
field in the gravitational multiplet $(g_{MN}, A_M, \Psi_M )$ form a
complex field $T$ with bosonic components $T = t + i A_5$. By ignoring
the one-loop corrections, it is well-known that the effective
lagrangian describing the radion with Scherk-Schwarz supersymmetry
breaking is of no-scale type with a constant superpotential
term~\cite{fkpz, dg,mp,kw}. The one-loop corrections produce a
deformation of the no-scale structure in the Kahler potential.  We
will use the relation between the 5D fundamental Planck scale $M_5$,
the 4D Planck mass $M_P$ and the fifth radius $R$ via
\be
R M_5^3 \ = \ M_P^2 \quad , \quad R = t M_5^{-1} \ , \label{o1}
\ee

We can parametrize the one-loop deformation of the no-scale structure
as~\footnote{In what follows we will use units where $M_P\equiv 1$.}
\be
K=- 3\log(T+\bar T)-\Delta K,\qquad \Delta K\equiv\frac{\Delta}{T+\bar T}
\ee
In the presence of a constant superpotential $W=\omega$ and in the large
radius limit we can relate $\Delta K$ with the one-loop effective potential
$V$ as
\be
V= e^K ( K^{T {\bar T}} |{D_T} W|^2 -3 |W|^2 ) =
\frac{|\omega|^2}{(T+\bar T)^2}\Delta^{\prime\prime}
\ee
where $\Delta^{\prime\prime}=\partial^2\Delta/\partial(T+\bar T)^2$.

We can now express the 4D effective supergravity Kahler potential in
the presence of massless hypers and massive hypers and vector
multiplets. For the case where Goldstone bosons come in
hypermultiplets it is given by as
\be
 \Delta= {1 \over (T+{\bar T})^2}
\left[ \alpha_0 + \alpha_1 (T + {\bar T}) e^{- \beta_1 (T + {\bar T})}
+ \alpha_2 e^{- \beta_2 (T + {\bar T})} \right] + \cdots\ ,
\label{o2}
\ee
where the dots represent subdominant terms in the large radius limit.
The scalar potential in the large radius limit $T+ {\bar T} \gg 1$ is
given by
\be
V  =
 {|\omega|^2 \over (T + {\bar T})^3}
\biggl[ {6 \alpha_0 \over (T + {\bar T})^3} + \alpha_1 \beta_1^2
e^{- \beta_1 ( T + {\bar T})} + {\alpha_2 \over (T + {\bar T})} \beta_2^2
e^{- \beta_2 ( T + {\bar T})} + \cdots \biggr] \ . \label{o3}
\ee
By using (\ref{o1}), we find
that the scalar potential (\ref{o3}) agrees with the leading contribution
to the vacuum energy (\ref{o01}) in the large radius approximation
for
\ba
&& \alpha_0 \sim {-\,2(N_h-2-N_v) \over \omega^2} \sum_{k=1}^{\infty}
{\sin^2 (\pi \omega k) \over k^5}
\quad , \quad \alpha_1 \sim N_H \quad , \quad
\alpha_2 \sim {-\,8 \sin^2 (\pi \omega) \over \omega^2 } N_V \ , \nonumber \\
&& \beta_1 \sim \pi M_H \quad , \quad  \beta_2 \sim \pi M_V \ . \label{o4}
\ea
The main reason in getting the possibility of having positive (de
Sitter) or zero vacuum energy is the oppositive sign contribution
of the hypers (coefficient $\alpha_1$) and of the vectors
(coefficient $\alpha_2$) in the one-loop vacuum energy (\ref{o3}).
Finally notice that the functional form of (\ref{o3}) is similar
to scalar potential in racetrack models of gaugino condensation
\cite{racetrack}.

For the case where Goldstone bosons absorbed by massive gauge
multiplets come in vector multiplets (\ref{o2})
should be replaced in the large radius limit by
\ba \Delta&=& {1 \over (T+{\bar T})^2} \left\{ \alpha_0 + \alpha_1 (T +
{\bar T}) e^{- \beta_1 (T + {\bar T})}\right. \nonumber\\ &+&
\left.\frac{\alpha_3}{(T+\bar T)^2}\left[3\,Li_7\left(e^{2
i\pi\omega+i\beta_3(T+\bar T)}\right)-3\,Li_7\left(e^{i\beta_3(T+\bar
T)}\right)+{\rm h.c.}  \right]\right\} ,
\label{o5}
\ea
where
\be
\alpha_3\sim -\frac{N_V}{\omega^2 M_V^2},\quad \beta_3\sim \pi M_V\ .
\ee
Notice that the oscillatory behaviour of the potential is transmitted
to the Kahler potential.

\section{Conclusions}

The goal of the present paper was to show that, generalizing
Ref.~\cite{pp} by including massive five-dimensional vector
multiplets together with massive 5D hypermultiplets of different
masses and combining them with supersymmetry breaking \'a la
Scherk-Schwarz, quantum loops generate a radion potential which
can stabilize it at a positive or zero vacuum energy. There are
two different ways of Higgsing a vector multiplet in 5D, by
absorbing either a hyper Goldstone, or by absorbing the fifth
component of another gauge field of appropriate parity. In an
orbifold compactification the second case asks for a non-abelian
gauge structure, generate a spontaneous breaking of the 5D Lorentz
invariance and is equivalent to the Hosotani mechanism of gauge
symmetry breaking. The four-dimensional description of this
phenomenon, valid at large radii, involves a one-loop modification
of the Kahler potential, which contains two (or more) exponential
terms coming with different signs. We would like to comment here
on the differences between the stabilisation presented here and
the one put forward in \cite{kklt} in the context of flux
compactifications.  The authors of \cite{kklt} use fluxes in order
to generate a constant in the superpotential, in combination with
nonperturbative effects like D3 instantons or gaugino condensation
on D7 branes in order to produce a nonperturbative superpotential
for the radion. A stabilisation at large radius in their case need
a certain amount of tuning between several fluxes in order to
generate a small constant in the superpotential. The present
mechanism does not need any particularly small numbers or
nonperturbative effects, whereas massive vector multiplets and
hypermultiplets are generally present in various string
compactifications. For a qualitatively different proposal of the
role of the massive vectors in string theory, see also
\cite{watson}.

There are several possible generalizations of our work. First of all,
a full 5D supergravity description of the two different realizations
of the Higgs phenomenon would be very useful for further studies and
for a more detailed analysis of the 4D supergravity description of the
stabilization mechanism of the Minkowski and de Sitter
solutions. Secondly, it would be interesting to combine, in the
framework of string models, the present stabilization with other
mechanisms of moduli stabilisation present in the literature in order
to find realistic models with all moduli stabilized.

\setcounter{footnote}{0}
\section*{Acknowledgments}
This work is supported in part by the CNRS PICS no. 2530 and 3059,
INTAS grant 03-51-6346, the RTN grants MRTN-CT-2004-503369,
MRTN-CT-2004-005104, by a European Union Excellence Grant,
MEXT-CT-2003-509661 and partly
by CICYT, Spain, under contracts FPA 2004-02015 and FPA 2002-00748 and
by the CICYT/IN2P3 grant IN2P3 05-19.  We are grateful to P. Fayet
and A. Pomarol for very useful discussions.  ED would like to acknowledge the
hospitality of the theory group of IFAE, Autonomous Univ.~of Barcelona
during the completion of this work.


\end{document}